\documentclass[
reprint,
superscriptaddress,
amsmath,amssymb,
aps,
prb,
floatfix,
]{revtex4-2}
\usepackage{float}
\pdfoutput=1
\usepackage{xcolor}
\usepackage{graphicx}
\usepackage{dcolumn}
\usepackage{bm}
\begin{document}

\title{Nucleation of magnetic textures in stripe domain bifurcations for reconfigurable domain wall racetracks}

\author{V.V. Fernández}
\affiliation{Depto. Física, Universidad de Oviedo, 33007 Oviedo, Spain.\\} 
\affiliation{CINN (CSIC–Universidad de Oviedo), El Entrego, Spain.\\}
\author{S. Ferrer}
\affiliation{ALBA Synchrotron, 08290 Cerdanyola del Vallès, Spain.\\}
\author{A. Hierro-Rodríguez}
\affiliation{Depto. Física, Universidad de Oviedo, 33007 Oviedo, Spain.\\} 
\affiliation{CINN (CSIC–Universidad de Oviedo), El Entrego, Spain.\\}
\author{M. Vélez}
\affiliation{Depto. Física, Universidad de Oviedo, 33007 Oviedo, Spain.\\} 
\affiliation{CINN (CSIC–Universidad de Oviedo), El Entrego, Spain.\\}

\date{\today}

\begin{abstract}
Within the racetrack memory paradigm, systems exploiting magnetic guiding potentials instead of geometrical ones, allow for enhancing the versatility of the final devices adding magnetic reconfigurable capabilities. Hard/soft magnetic multilayers with stripe domain configurations fulfill these requirements. In these systems, the topology of the generated textures that would act as information carriers, is strongly conditioned by the stripe lattice configuration. Micromagnetic simulations have been used to study the magnetization reversal process in NdCo$_5$/Py reconfigurable racetracks. By using skyrmionic charges and magnetic vorticity lines, the topological transformations controlling the nucleation of vortices, antivortices, Bloch lines and Bloch points has been analyzed. It has been shown that magnetic topological charge exchanges between textures rule the formation of vortex/antivortex pairs with opposite polarities, key for the guided propagation through the stripe pattern.  

\end{abstract}

\maketitle

\section{\label{sec:Intro} Introduction}

The design of spintronic devices such as racetrack memories \cite{DWracetrack,Parkin,skyrmion.racetrack} or domain wall logic units \cite{Dedalo,DWlogic,Dedalo2} requires a good control of the nucleation and propagation mechanisms of magnetic textures in thin-films. The topological restrictions of magnetic textures in two-dimensional (2D) confined geometries \cite{vortex.frac, DW.topo, VAV.square} and chiral interactions such as Dzyaloshinskii–Moriya \cite{skyrmion.racetrack} provide robust restrictions in the configuration of propagating spin textures. 
In magnetic multilayers, exchange and dipolar couplings across the thickness offer an additional toolbox for designing reconfigurable devices with guiding potentials of magnetic origin \cite{reconfig2, APL2017, Italiano,dipolar, Li2021, reconfig1, PhysRevApplied.23.014023} and three-dimensional (3D) magnetic memories \cite{Cowburn13}. 

Magnetic interactions across the thickness of the sample can also give rise to a variety of 3D magnetic textures in bulk materials and nanostructures, such as skyrmion and antiskyrmion tubes \cite{skyrmion.tube, antiskyrmion.tube}, curved vortex strings in magnetic nanocaps \cite{vortex.tube}, or fractional skyrmion tubes in magnetic nanohelices \cite{frac.sky}. These 3D magnetization patterns often contain Bloch point singularities \cite{BP.chiral, BP.Donnelly, BP.dipole} that are nucleated during magnetization reversal to meet the topological restrictions of 3D magnetic samples \cite{BP.topology}. A typical example could be the reversal of core polarity in a magnetic vortex in a nanodot, which reverses its skyrmionic charge ($Q_z$) from $Q_z=+1/2$ to $Q_z=-1/2$ \cite{GOBEL20211}) and requires the propagation of a Bloch point across the vortex core line for the topological charge increment \cite{BP.vortexreversal}. In this context, magnetic vorticity ($ \mathbf{\Omega}$) and emergent field  ($\mathbf{B^e}$) lines are valuable tools to describe the interactions between the different textures present in a 3D magnetization configuration \cite{omega.claire,NC2020,BP.dipole}. 

\begin{figure}[ht]
    \centering
    \includegraphics[width = 1\linewidth]{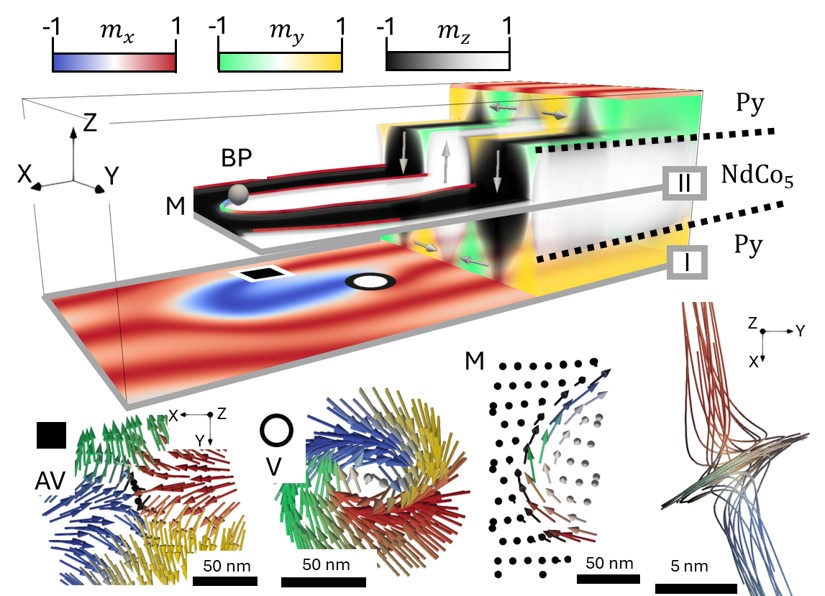}
        \caption{\label{fig:sketch}Sketch of the magnetic configuration in a Py/NdCo$_5$/Py multilayer and its relevant textures at different layers. Each component of the unit magnetization vector is shown with a different color scale: $m_x$, blue to red; $m_y$, green to yellow; and $m_z$, black to white. Grey arrows are a guide to the eye to understand the magnetization configuration throughout the sample thickness. Planes I and II correspond to the bottom Py surface and the central NdCo$_5$ layer, respectively. In plane I two topological textures can be seen: a vortex (V), marked with a white circle, and an antivortex (AV), black square. Symbol color indicates $m_z$ orientation at V/AV core: white for $+m_z$ polarity and black for $-m_z$ polarity. Plane II contains a meron-like texture (M) and a Bloch Point (BP). The insets at the bottom show the magnetic configuration of these four textures.}
\end{figure}

Recently, a reconfigurable domain wall racetrack has been proposed based on exchange-coupled hard/soft magnetic layers \cite{PhysRevApplied.23.014023} as sketched in Fig.\ref{fig:sketch} (arrows and color codes indicate the unit magnetization vector $\textbf{m}=(m_x,m_y,m_z)$ and the sign of each magnetization component $m_i$). There is a central hard magnetic layer with weak perpendicular magnetic anisotropy (e.g. NdCo$_5$ in Fig. \ref{fig:sketch}), hosting a periodic pattern of parallel up/down stripe domains that act as guiding channels for the magnetic textures that propagate under the effect of current pulses at the outer soft magnetic layers (e.g., Permalloy (Py) in Fig.\ref{fig:sketch}). At remanence, the in-plane orientation of the propagation paths (i.e. the stripe domains) is defined by the last in-plane saturating magnetic field (along the positive $x$-axis in Fig.\ref{fig:sketch}), so that it can be easily determined playing with the magnetic history of the system. In-plane magnetization reversal of the parallel stripe domain pattern is subject to strong topological restrictions \cite{topological}, so that stripe domain bifurcations become the preferred loci for nucleation of reversed domains and magnetic textures \cite{APL2017,PRB2017}. 

The typical initial state of the reconfigurable racetrack contains a vortex (V)-antivortex (AV) pair in the soft Py layer (plane I in Fig.\ref{fig:sketch}) that bounds a small $-m_x$ domain at the bifurcation core. In the central NdCo$_5$ layer (plane II in Fig.\ref{fig:sketch}), the bifurcation core ($+m_z$ white stripe end-point) hosts a meron-like texture (M) and a Bloch point (BP) singularity that propagates along the boundary between $+m_z/-m_z$ stripes \cite{NC2015,NC2020}. The exchange and magnetostatic coupling between the NdCo$_5$ and Py layer provides both the magnetic guiding potential that defines the geometry of the racetrack and a bias field that controls the propagation direction of the textures within the racetrack, resulting in a memory effect \cite{PhysRevApplied.23.014023}. However, a good understanding of the mechanisms that generate each of these textures, needed to control their injection in the reconfigurable racetrack, is still lacking.

In this work, the evolution of the magnetic configuration at the bifurcation core during nucleation of the initial $-m_x$ reversed domain has been studied by micromagnetic simulations, and the different magnetic textures at the NdCo$_5$ and Py layers have been tracked by their skyrmionic charges. Magnetic vorticity lines have allowed us to visualize the role of the curved $+m_z/-m_z$ domain wall in the bifurcation core on the interactions between the different spin textures across the multilayer thickness. It is found that the generation of a pair of Bloch points with opposite topological charges  at the NdCo/Py interface is an essential ingredient to obtain the V/AV pairs of opposite polarities that propagate along parallel stripes within the reconfigurable racetracks.

\section{\label{sec:Calc} Micromagnetic characterization of spin textures at stripe domain bifurcations}

\subsection{\label{sec:Micro} Micromagnetic simulations}
\begin{figure}[ht]
    \centering
    \includegraphics[width = 1\linewidth]{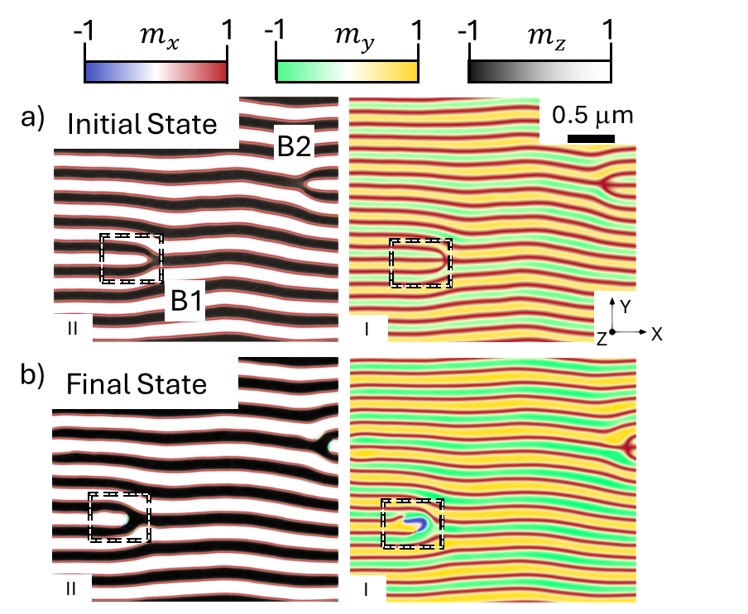}
        \caption{\label{fig:setup} Evolution of the magnetic configuration of the 40 nm Py/80 nm NdCo$_5$/40 nm Py multilayer along the nucleation process of magnetic textures at bifurcations $B1$ and $B2$: (a) Initial state $\mu_0H_x=30$ mT and (b) Final state $\mu_0H_x=0$ mT. Plane I corresponds to the bottom surface at Py layer and plane II to the central plane at NdCo$_5$ layer. Dotted rectangles indicate the region of interest to analyze the nucleation of magnetic textures at the core of bifurcation $B1$.}
\end{figure}

Micromagnetic simulations of a 40 nm Py/80 nm NdCo$_5$/40 nm Py multilayer have been performed with Mumax$^3$ \cite{Mumax} with magnetic parameters typical of the selected materials \cite{PhysRevApplied.23.014023}. In the NdCo$_5$ layer, the out-of-plane anisotropy is $K_z($NdCo$)=1.4\times 10^5$ J/m$^3$ and in-plane anisotropy $K_x($NdCo$)=4\times 10^3$ J/m$^3$ with the easy axis oriented along the $x$ direction. In the Py layer, the out-of-plane anisotropy is $K_z($Py$)=1.1 \times 10^4$ J/m$^3$ and in-plane anisotropy $K_x($Py$)= 850$ J/m$^3$ also with the easy axis along $x$. Uniform saturation magnetization $M_S=8\times 10^5$ A/m and exchange stiffness $A=1.3\times 10^{-11}$ J/m have been used for both materials. The simulation volume is $768 a_x \times 512 a_y\times 40 a_z$ with $a_x=a_z= 4$ nm and periodic boundary conditions along the $x$ and $y$ directions. The initial magnetic state was prepared in a three-step process, in a similar way as previously reported \cite{PhysRevApplied.23.014023}: first, the homogeneous stripe domain pattern is optimized (total energy minimization) at remanence as a function of the lateral discretization scale $a_y$ (with stripe domains oriented along $x$, positive $m_x$ in the full simulation volume and period $\Lambda=512 a_y/9$) and the resulting $a_y^{opt}=5.16$ nm is used in the subsequent simulations; second, stripe domain bifurcations are created at predefined locations in the system, to act as nucleation sites for spin textures; third, the system is relaxed under an in plane magnetic field $\mu_0H_x=30$ mT, large enough to ensure positive $m_x$ in the full simulation volume including bifurcation cores (see Fig. \ref{fig:setup}(a)). Finally, the applied field is removed and the evolution of the magnetic configuration of the system is monitored until a relaxed remanent state is reached (see Fig. \ref{fig:setup}(b)). This process ensures that the overall stripe pattern stays constant (nine stripe periods with two bifurcations within the simulated volume) and that the relevant changes in the magnetic configuration occur at the bifurcation cores. 

It should be noted that, due to symmetry considerations, the different possible types of bifurcation in a parallel stripe domain pattern can be reduced to only two \cite{PRB2017}, corresponding to bifurcations $B1$ and $B2$ in Fig. \ref{fig:setup}. Magnetostatic interactions across thickness break the symmetry between multilayer top / bottom surfaces, favoring nucleation of the $-m_x$ domains at the bottom surface for $B1$ and at the top surface for $B2$ \cite{PRB2017}. The nucleation process is essentially the same in both cases, so we will focus on the changes in magnetic configuration at the core of the bifurcation $B1$ (see dotted rectangles in Fig. \ref{fig:setup}). 

\subsection{\label{sec:Topo} 3D spin textures and skyrmionic charges}

Spin textures in thin magnetic layers and magnetic nanostructures are generally characterized by their skyrmionic charge $Q_z$. In high symmetry configurations, for example, a magnetic vortex in a thin disk, $Q_z$ can be estimated from simple parameters such as polarity ($p=\pm 1$ depending on the sign of $m_z$ at the core) and vorticity ($n=\pm1$ for a V or an AV, respectively) as $Q_z=\frac{1}{2} p\cdot n$ \cite{GOBEL20211}. For spin textures in extended films, as is the case here, a more accurate calculation of the total skyrmionic charge $Q_z$ of each texture can be obtained from the integral of the skyrmionic charge density $q_z$ \cite{GOBEL20211, BP.chiral, BP.dipole} as

\begin{equation}
Q_z=\iint q_z dxdy= \frac{1}{4\pi}\iint \mathbf{m}\cdot \left[\frac{\partial \mathbf{m}}{\partial x} \times \frac{\partial \mathbf{m}}{\partial y}\right] dxdy.
\end{equation}

\begin{figure}[ht]
    \centering
    \includegraphics[width = 1\linewidth]{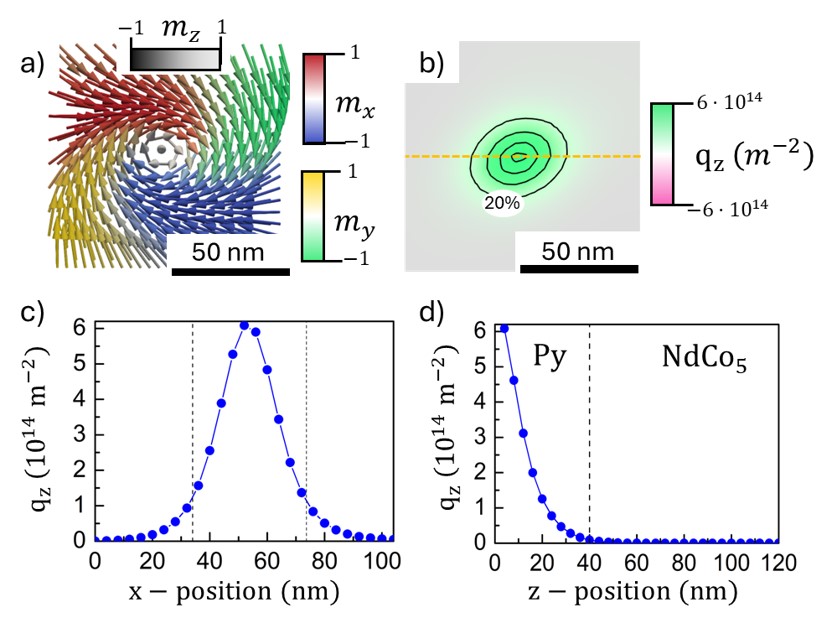}
        \caption{\label{fig:vortex} Description of the vortex present in the bottom surface of the Py layer at the final state ($\mu_0H_x=0$ mT). (a) Vector magnetic configuration. (b) Map of skyrmionic charge density ($q_z$) around the V. Black contour lines enclose the areas where $q_z$ is 20$\%$, 40$\%$, 60$\%$ and 80$\%$ of its maximum value $q_z^{max}$, respectively. (c) $q_z$ profile across the V core, following the dotted orange line in (b). The vertical dotted lines mark the position where $q_z = 0.2 q_z^{max}$. (d) $q_z$ profile throughout the thickness of the Py and NdCo$_5$ layers, calculated at the V core. The vertical dotted line marks the NdCo/Py interface.}
\end{figure}

Figure \ref{fig:vortex} shows the magnetic configuration (and its corresponding $q_z$ map calculated with Eq. (1)) for a magnetic vortex with positive polarity $p=+1$, located close to bifurcation $B1$ on the bottom multilayer surface. The total skyrmionic charge of this texture is $Q_z(V)=0.41$ calculated from the integral of $q_z$ over a circular region of $r = 40$ nm centered at the vortex. This value, clearly below the ideal $Q_z=1/2$ for a vortex embedded in a film with in-plane magnetization \cite{GOBEL20211}, is typical for vortices in the stripe domain pattern due to the non-trivial magnetization background \cite{PhysRevApplied.23.014023}. As shown in Fig. \ref{fig:vortex} (b), the vortex $q_z$ is highly concentrated: it reaches a maximum value $q_z^{max}\approx 6\times 10^{14}$ m$^{-2}$ in the vortex core and then decreases to $0.2 q_z^{max}= 1.2\times 10^{14}$ m$^{-2}$ just 20 nm from it (see $q_z$ \textit{vs.} $x$ profile across the vortex core in Fig. \ref{fig:vortex} (c)). In fact, the reduced skyrmionic charge calculated only within the contour of $0.2q_z^{max}$ in Fig. \ref{fig:vortex} (b)) is $Q_z^*(V)=0.28 \approx 0.56\times Q_z(V)$, suggesting that these high $q_z$ regions can be used to identify the position of each spin texture in the multilayer \cite{BP.dipole,vortex.tube}. $q_z$ is also the highest close to the multilayer surface (see the steep decrease of $q_z$ \textit{vs.} $z$ at the vortex core shown in Fig. \ref{fig:vortex} (d)) indicating that this vortex texture is mainly confined within the bottom Py layer. Therefore, in the subsequent analysis of the magnetic configuration in the bifurcation core, $q_z$ isosurfaces will be used to locate the relevant spin textures within the simulated volume using the color code: green isosurface ($q_z^{+}=0.2q_z^{max}=+1.2\times 10^{14}$ m$^{-2}$), pink isosurface ($q_z^{-}=-0.2q_z^{max}= -1.2\times 10^{14}$ m$^{-2}$). 

\begin{figure}[ht]
    \centering
    \includegraphics[width = 1\linewidth]{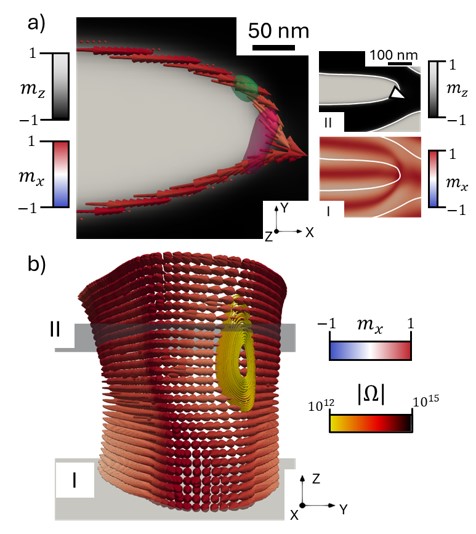}
        \caption{\label{fig:curvedDW} Magnetization configuration during the first simulation steps, once $\mu_0H_x=0$ mT. (a) Magnetization in the central NdCo$_5$ plane (plane II). $m_z$ component is shown in black and white, while $m_x$ is shown with arrows and in a blue/red color scale. The green/pink areas correspond to isosurfaces of $q_z= \pm 1.2\times 10^{14}$ m$^{-2}$ respectively. The insets in (a) will be used during the article as guides to show the magnetic configuration on planes I and II in each step. The white lines delimit the $m_z$ = 0 DW at plane II in both insets. White triangle marks the BL location at plane II. (b) 3D view of the magnetization in the $+m_z/-m_z$ DW that delimits the central white stripe domain (calculated as the $m_z$ = 0 surface across the thickness). When the magnetization reversal begins this DW curves inward and a bump can be seen close to plane II. This curvature gives rise to the magnetic vorticity ($\mathbf{\Omega}$) represented as rings in a yellow/black scale.}
\end{figure}

Figure \ref{fig:curvedDW} shows an example of magnetic textures within the NdCo layer identified by two high $|q_z|$ regions close to the bifurcation core (marked by the green and pink isosurfaces on the domain wall (DW) between the $+m_z/-m_z$ domains in Fig. \ref{fig:curvedDW}(a)). Near the center of the multilayer (plane II), this DW has a clear Bloch character with in-plane magnetization approximately along $+m_x$, i. e. parallel to the DW plane, at the straight sections bounding the central white stripe of the bifurcation. Then, at the white stripe endpoint, both $+m_x$ branches meet head-to-head, creating a Bloch line (BL) texture \cite{huber} (see white triangle at inset II of Fig. \ref{fig:curvedDW}(a)). This BL displays a high negative $q_z$ density (pink isosurface in Fig. \ref{fig:curvedDW}(a)) and therefore a negative skyrmionic charge $Q_z^*(BL)<0$. Near the BL, there is a region with high positive $q_z$ values (green isosurface in Fig. \ref{fig:curvedDW}(a)) corresponding to the curved DW section between the $+m_z/-m_z$ domains. This texture with $Q_z^*(M)>0$ will be labeled as meron-like (M) since the curved stripe end points can reach skyrmionic charges up to $\pm0.5$ \cite{meron.ezawa,NC2020}.

To understand the connections between the different textures observed (vortices, antivortices, meron-like, Bloch lines and Bloch points), we need to go beyond skyrmionic charges and consider the full magnetic vorticity $\mathbf{\Omega}$ or its corresponding emergent field $\mathbf{B^e}=\hbar \mathbf{\Omega}$ \cite{KOMINEAS199681,NC2020, omega.claire} that provide a more general description of the topological properties of the magnetization in 3D:  
\begin{equation}
B^e_i=\hbar \Omega_i=\frac{\hbar}{2} \epsilon_{ijk}  \mathbf{m}\cdot \left[\partial_j \mathbf{m} \times \partial_k \mathbf{m}\right]. 
\end{equation}
In this framework, the skyrmionic charge $Q_z$ is simply the flux of $\mathbf{\Omega}$ across the $xy$ plane and $q_z$ the projection of $\mathbf{\Omega}$ normal to the multilayer surface. 
Fig. \ref{fig:curvedDW}(b) shows the $\mathbf{\Omega}$ lines associated with the BL and M textures on top of the $m_z=0$ DW surface in 3D. The DW surrounds the white stripe endpoint with an inward bump across the multilayer thickness, marked by a set of rings of $\mathbf{\Omega}$ lines. The upward / downward intersections of $\mathbf{\Omega}$ lines across plane II give rise to the high $|q_z|$ regions of the M / BL textures with opposite $Q_z^*$ values observed in Fig. \ref{fig:curvedDW}(a)). 
Thus, both $\mathbf{\Omega}$ lines and $q_z$ isosurfaces will be used to track the connections between the spin textures at the different layers within the stripe bifurcation core.

\section{\label{sec:RyD} Analysis of magnetic texture nucleation at bifurcation core}

\begin{figure*}[ht]
    \centering
    \includegraphics[width = 0.7\linewidth]{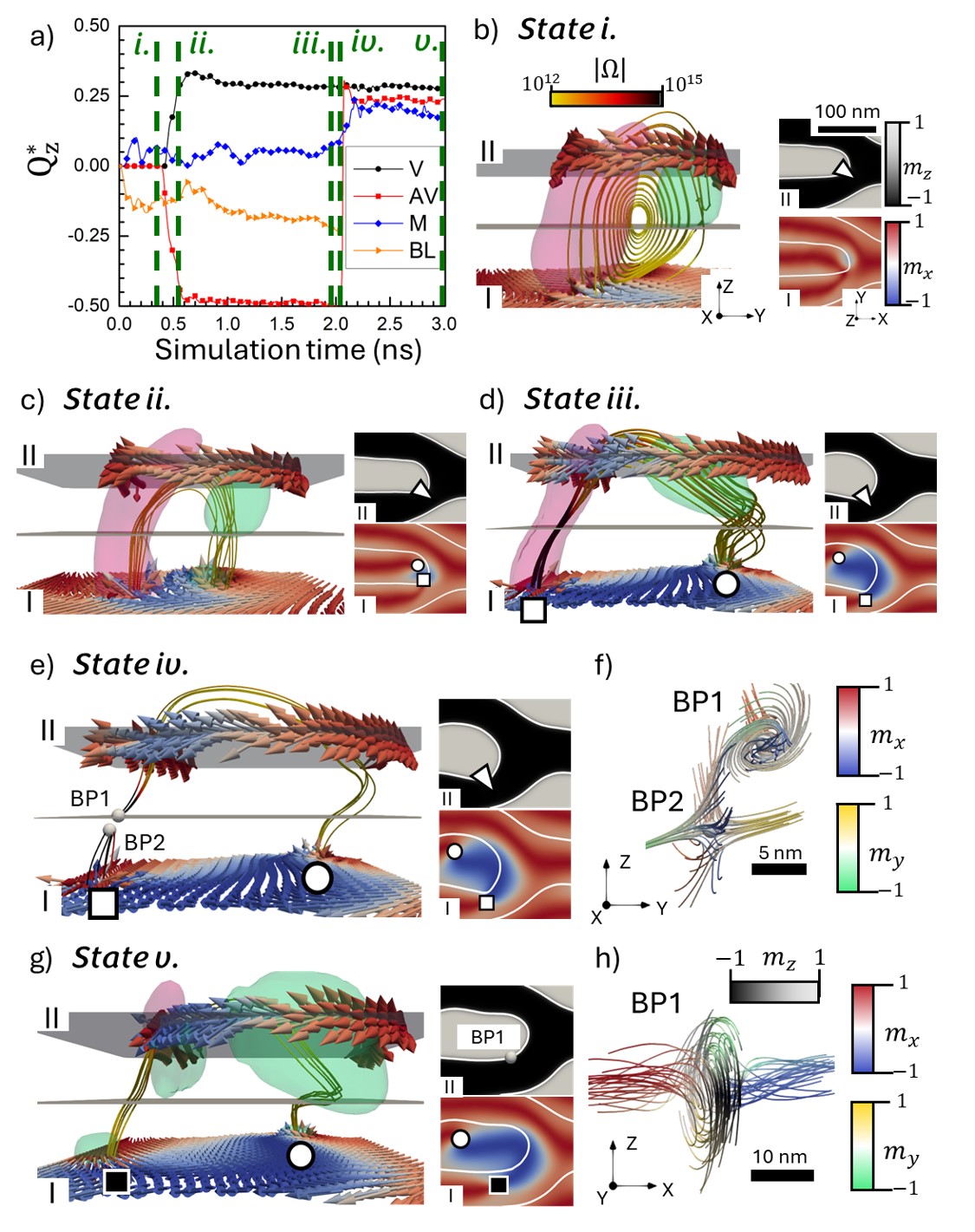}
        \caption{\label{fig:grande} (a) Reduced skyrmionic charge $Q_z^*$, integrated within the $q_z^{+}$/$q_z^{-}$ contours as a function of simulation time, along the nucleation
process of magnetic textures at bifurcation B1: the vortex at plane I ($\textcolor{black}{\bullet}$), the antivortex at plane I ($\textcolor{red}{\blacksquare}$), the meron-like texture at plane II ($\textcolor{blue}{\blacklozenge}$) and the Bloch line at plane II ($\textcolor{orange}{\blacktriangleright}$). Snapshots of magnetic configuration at \textbf{States}: (b) \textbf{\textit{i}}, $t=0.35$ ns, with only BL (white triangle) and M textures; (c) \textbf{\textit{ii}}, $t=0.55$ ns, with the initial V/AV pair (white circle/square) of $+m_z$ polarity located below the $+m_z$ stripe; (d) \textbf{\textit{iii}}, $t=1.95$ ns, with the $+m_z$ polarity AV shifted towards the $-m_z$ stripe; (e) \textbf{\textit{iv}}, $t=2.03$ ns, with Bloch points BP1-BP2 (grey spheres) near the NdCo/Py interface (stream lines of the magnetic configuration around BP1 and BP2 are shown in (f)); (g) \textbf{\textit{v}}, final state obtained when $t = 3$ ns, with the desired configuration for the reconfigurable DW racetrack: opposite polarity V/AV textures located under stripe domains with matching $m_z$ sign and a BP at plane II within the $+m_z/-m_z$ DW (Magnetization streamlines of the magnetic configuration of BP1 are shown in (h)). The magnetization at the bottom surface of Py (plane I) and in the central NdCo$_5$ layer (plane II) are represented with arrows and the $m_x$ component is colored in a blue/red scale. The NdCo$_5$/Py interface is shown with a grey plane. The $q_z^{+}$/$q_z^{-}$ isosurfaces are shown in green/pink, respectively.  Note that, the $q_z$ dipole (pink/green isosurfaces in contact) that apperas in (g) corresponds to BP1. The magnetic vorticity ($\mathbf{\Omega}$) lines are represented in a yellow/black scale. The insets on planes I and II are guides to show the position of the topological textures in each state relative to the $+m_z/-m_z$ DW at plane II (white contours).}
\end{figure*}

Figure \ref{fig:grande}(a) describes the evolution of the reduced skyrmionic charges for the different magnetic textures that appear during nucleation of a reversed $-m_x$ domain at the core of bifurcation $B1$. Snapshots of the magnetic configuration at \textbf{States \textit{i}}-\textbf{\textit{v}} are shown in Figs.  \ref{fig:grande}(a-g) with a plot of the $\mathbf{\Omega}$ lines that connect the different textures in the sample to analyze their topological relationships. 

In the initial state (as shown in Fig. \ref{fig:setup}(a)), the multilayer is relaxed under the effect of an applied $\mu_0H_x=30$ mT and no textures are detected in the system ($t=0$). As soon as this applied field is removed and the multilayer is allowed to relax toward the remanent state, the magnetization within the bifurcation core starts to rotate away from $+m_x$ driven by two main factors: a) the out-of-plane anisotropy $K_z($NdCo$)$ that favors $\pm m_z$ components at the NdCo layer and b) the magnetostatic field that circulates from the $-m_z$ to the $+m_z$ domain and favors $-m_x$ components at the bottom of the multilayer \cite{PRB2017}. 

Within $0<t<0.4$ ns, the increasing $\pm m_z$ amplitude near the curved DW at the white stripe endpoint creates two high $|q_z|$ regions of opposite signs in the center of the NdCo$_5$ layer (plane II). They correspond to the Bloch line ($Q_z^*(BL)<0$) and the meron-like texture ($Q_z^*(M)>0$) (see, e.g., \textbf{State \textit{i}} at $t=0.35$ ns in Fig. \ref{fig:grande}(b)). A small $-m_x$ domain is observed in plane I, created by the closure magnetostatic field that is largest at the bottom of the multilayer. $\mathbf{\Omega}$ lines form closed rings that spread over a wide area of the white stripe endpoint, indicating a complex 3D curved shape of the $+m_z/-m_z$ DW. Their downward branch (indicated by a broad tubular pink $q_z^-$ contour) corresponds to a vertical BL located on the DW (see the white triangle in inset II in Fig. \ref{fig:grande}(b)). The upward $\mathbf{\Omega}$ ring branch (tubular green $q_z^+$ contour) is the M texture. $Q_z^*(M)\approx-Q_z^*(BL)\approx 0.15$  correspond to the flux created at the intersection of the upward/downward $\mathbf{\Omega}$ ring branches with plane II.

Up to $t\approx 0.4$ ns, there is a gradual expansion of the $\mathbf{\Omega}$ rings, with an increase in the length of the BL toward the bottom Py layer until eventually $\mathbf{\Omega}$ rings of sufficient intensity reach the multilayer surface near the $-m_x$ domain. Then, between $t=0.4-0.5$ ns (see Fig. \ref{fig:grande}(a), there is a fast increase in $\mathbf{\Omega}$ flux through plane I that marks the nucleation of the V/AV textures on the bottom surface of the multilayer with $Q_z^*(V)$ and $Q_z^*(AV)$ of opposite signs (calculated on plane I). Figure \ref{fig:grande}(c) (\textbf{State \textit{ii}}, at $t=0.55$ ns) shows that the point of emergence of the pink $q_z^-$ contour corresponds to the AV, strongly coupled to the vertical BL. On the other hand, upward $\mathbf{\Omega}$ flux is more broadly spread. This results in lower positive $q_z$ values and a discontinuous green $q_z^+$ contour within the Py layer, indicating a weaker coupling between the M and V textures. Both V and AV are located just below the NdCo white $+m_z$ stripe domain and, thus, display $+m_z$ polarity at their cores (see the white circle and the white square at inset I in Fig. \ref{fig:grande}(c)).

Subsequent expansion of the Py $-m_x$ domain occurs by propagation of the V / AV textures away from the bifurcation core, as shown in Fig. \ref{fig:grande}(d) corresponding to \textbf{State \textit{iii}} at $t=1.95$ ns. The loosely bound vortex moves easily along the NdCo $+m_z$ stripe, increasing the length of the Py $-m_x$ domain. The AV makes a short lateral displacement towards the NdCo $-m_z$ stripe, which increases the width of the Py $-m_x$ domain. In this way, the Py $-m_x$ domain takes advantage of the high $-H_x$ intensity of the closure magnetostatic field just below the $+m_z/-m_z$ boundary at the bifurcation core (see insets I\&II of Fig. \ref{fig:grande}(d)). This lateral AV displacement pulls the BL string from its lower end and distorts the entire $+m_z/-m_z$ DW across the multilayer thickness. The BL adopts a twisted shape with a very high $\mathbf{\Omega}$ intensity at the boundary between the NdCo and Py layers corresponding to the region of highest DW curvature across the thickness (see the Supplementary Video \cite{suppl.movie}).

Up to \textbf{State \textit{iii}}, the evolution of the magnetic configuration of the multilayer can be understood as a continuous deformation of the DW at the bifurcation core, in which the relevant spin textures appear on the high $\mathbf{\Omega}$ regions within the sample volume (BL \& M) or at their emergence points at the sample surface (V \& AV). However, this continuity has led the system to a highly unstable configuration with an AV of $+m_z$ polarity at the Py layer located below the $-m_z$ NdCo stripe, i.e. with a section of the $+m_z/-m_z$ DW with head-to-head configuration across the thickness. At \textbf{State \textit{iv}}, at $t=2.03$ ns (Fig. \ref{fig:grande}(e)), we observe that a pair of Bloch points appears just at the NdCo/Py interface, breaking the BL string at the point of highest $\mathbf{\Omega}$ intensity. As shown in Fig. \ref{fig:grande}(f), Bloch point 1 (BP1) presents a circulating configuration with head-to-head $+m_x/-m_x$ reversal along the symmetry axis surrounded by a circulating vortex in $m_y/m_z$, corresponding to a circulating BP of negative topological charge $Q(BP1)=-1$ \cite{BP.dipole,BP.hyperbolic}. Bloch point 2 (BP2) presents a hyperbolic configuration with tail-to-tail $+m_y/-m_y$ reversal along the symmetry axis surrounded by an inward radial vortex in $m_x/m_z$, corresponding to a hyperbolic BP of positive topological charge $Q(BP2)=+1$ \cite{BP.hyperbolic}. Nucleation of the BP1-BP2 dipole conserves the total topological charge in the multilayer (see \textbf{States \textit{iii}} and \textbf{\textit{iv}} at Fig.\ref{fig:grande}(a)) and provides a mechanism to reduce the energy of the deformed $+m_z/-m_z$ DW. 

As the simulation progresses, the DW retracts into a straight configuration across the thickness, the two BL branches shrink until the BL disappears and there is a sharp upward step in the values of $Q_z^*(M)$ and $Q_z^*(AV)$ at $t\approx 2$ ns (see Fig.\ref{fig:grande}(a)). This large $\Delta Q_z^*(AV)$ is caused by the propagation of BP2 towards the Py surface where it recombines with the AV of positive $+m_z$ polarity. The skyrmionic charge of the resulting texture can be estimated as $Q(AV)_{final}=Q(AV)_{iv}+Q(BP2)=-0.5+1=+0.5$, qualitatively in agreement with $Q^*(AV)_{final}\approx +0.25$ (see Fig. \ref{fig:grande}(a)). 
An important consequence of the sign change of $Q(AV)$ is that the AV polarity switches to $-m_z$. Thus, in the final stable remanent state, AV polarity matches the surrounding $-m_z$ stripe domain (see black square in the inset of \ref{fig:grande}(g)). A single Bloch point remains in the multilayer: BP1 located in the center of the NdCo layer (plane II) with its symmetry axis aligned with the boundary in between $+m_z/-m_z$ stripe domains.
Once the BL disappears, the intensity of $\mathbf{\Omega}$  lines joining the AV with plane II drops by two orders of magnitude. This indicates a weaker coupling between the textures in the NdCo layer (BP \&  M) and the texture in the Py layer (V \& AV), so that they will be able to propagate independently.  

Therefore, the multilayer has finally reached the desired configuration for texture injection in a reconfigurable racetrack \cite{PhysRevApplied.23.014023} with a V with $Q(V)>0$ and positive polarity placed under a NdCo $+m_z$ stripe and an AV with $Q(AV)>0$ and negative polarity located under a NdCo $-m_z$ stripe. Lateral displacements of these textures towards a stripe of opposite $m_z$ sign are hindered by either the steep increase in magnetostatic energy or the need to nucleate a BP to reverse the texture polarity. Thus, the combination of magnetostatic and topological mechanisms in this magnetic configuration provides the very effective guiding of magnetic textures observed both in pulsed field \cite{APL2017} and pulsed current \cite{PhysRevApplied.23.014023} experiments.

\section{\label{sec:Conclusions} Conclusions}
The nucleation of magnetic textures at a bifurcation core in the stripe domain pattern has been studied by micromagnetic simulations during the initial stages of $-m_x$ magnetization reversal to understand the initial magnetic state on NdCo/Py reconfigurable racetracks.

Magnetic vorticity rings mark the deformations in the curved $+m_z/-m_z$ DW surrounding the white stripe endpoint and generate the initial high $q_z$ regions at the central NdCo plane that correspond to a BL and a M texture of opposite skyrmionic charges. As the magnetization reversal progresses, driven by the closure magnetostatic field of the $+m_z$ and $-m_z$ domains, high intensity $\mathbf{\Omega}$ lines begin to cross the Py surface, resulting in nucleation of a V/AV pair with opposite $Q_z$. The lateral propagation of the AV away from the bifurcation core twists the BL across the thickness and creates an unstable configuration with $+m_z/-m_z$ reversals across the thickness. Eventually, a Bloch point dipole generated at the NdCo/Py interface provides the essential mechanism to reverse AV polarity and to obtain the remanent sate with V/AV pairs of opposite polarities needed for guided propagation along parallel stripes within the reconfigurable racetracks. \cite{open.data}

\acknowledgments
VVF acknowledges support from the Severo Ochoa Predoctoral Fellowship Program (nos. PA-22-BP21-124). This work has been supported by Spanish MCIN/AEI/10.13039/501100011033/ FEDER,UE under grant PID2022-136784NB and by Agencia SEKUENS (Asturias) under grant UONANO IDE/2024/000678 with the support of FEDER funds. 

\bibliography{apssamp}

\end{document}